\documentclass[useAMS,usenatbib, letters]{mnras}

\usepackage{graphicx}   
\usepackage[english]{babel}

\newcommand{\SII}{[S~{\sc ii}]\ }
\newcommand{\OIII}{[O~{\sc iii}]\ }

\newcommand{\HII}{H~{\sc ii}\ }
\newcommand{\HeII}{He~{\sc ii}\ }

\newcommand{\Ha}{H$\alpha$\ }
\newcommand{\kms}{\,\mbox{km}\,\mbox{s}^{-1}}
\newcommand{\ergs}{\,\mbox{erg}\,\mbox{s}^{-1}}
\newcommand{\HST}{\textit{HST}\ }
\newcommand{\SIIHa}{I([S~{\sc ii}])/I(H$\alpha$)}

\newcommand{\be}{\begin{equation}}
\newcommand{\ee}{\end{equation}}

\def \gtsima{$\, \buildrel > \over \sim \,$}
\def \ltsima{$\, \buildrel < \over \sim \,$}
\def \simgt{\lower.5ex\hbox{\gtsima}}
\def \simlt{\lower.5ex\hbox{\ltsima}}
\defcitealias{Egorov16}{Paper~I}

\begin{document}

\title[Ionized gas around ULX HoII X-1.] {The ultraluminous X-ray source HoII X-1: kinematic evidence of its escape from the cluster. }

\author[Egorov et al.]{
   Oleg V.~Egorov$^{1}$\thanks{E-mail: egorov@sai.msu.ru},
   Tatiana A.~Lozinskaya$^{1}$,
   and Alexei V.~Moiseev$^{1,2}$ \\
$^{1}$ Sternberg Astronomical Institute, Lomonosov Moscow State University, 
        Universitetsky pr. 13, Moscow 119234, Russia
       \\
  $^{2}$ Special Astrophysical Observatory, Russian Academy of Sciences, Nizhnii Arkhyz 369167, Russia
   \\
      }

\date{Accepted 2016 Month 00. Received 2016 Month 00; in original
form 2016 Month 00}

\pagerange{\pageref{firstpage}--\pageref{lastpage}} \pubyear{2016}

\maketitle

\label{firstpage}

\begin{abstract}

We analyse the structure and kinematics of ionized gas in the vicinity of the ultraluminous X-ray source (ULX) HoII X-1 in the Holmberg~II galaxy using observational data obtained with a scanning Fabry--Perot interferometer in the H$\alpha$, \SII and \OIII emission lines at the Russian 6-m telescope. Decomposition of the line profiles allows us to identify the broad component of emission lines caused by the ULX action. We found evidence of an expanding superbubble around the young star cluster located in the studied region. We conclude that the blue-shifted `arc' around the ULX observed in the line-of-sight velocity field may correspond to a bow shock caused by the ULX movement from that nearby young star cluster. If this interpretation is correct, it will be the first kinematic evidence of ULX's escape from their parent star clusters.

\end{abstract}


\section{Introduction}

Ultraluminous X-ray sources (ULXs) are unique off-nucleus
point-like objects, which luminosities ($10^{39} - 10^{41}\
\mathrm{erg\ s^{-1}}$) exceed the Eddington limit. Their nature is
under debates over twenty years. These objects are often
considered as the intermediate mass black hole (IMBH) candidates
\citep[e.g.,][]{colbert99, miller03}, while the recent studies
support an idea of the supercritical accretion discs around the
stellar mass black holes as the source of anisotropic X-ray
radiation of ULXs \citep[e.g.,][]{begelman02, fabrika15,
roberts16}.

Studies of extragalactic ULXs revealed their association with
young star clusters \citep*{zezas02, abolmasov07, swartz09}. \citet{poutanen13}
showed that all bright X-ray sources (including 6 ULXs) in
Antennae galaxies are  significantly associated with young
clusters, but they are located outside these clusters. Together
with similar findings by \cite{kaaret04b} for X-ray binaries
of lower luminosities, this result allows one to assume that the
offset from star clusters is a common property of both ULXs
and less bright X-ray binaries. \citeauthor{poutanen13}
concluded that large displacements (up to 300 pc) are most
probably caused by their fast ($\sim80 \kms$)
ejection from clusters  due to close encounters
\citep*[e.g.,][]{mapelli11, goswami14}. Despite the fact that
observations reveal displacement of ULXs, no direct evidence
of their motion outward from star clusters has been found so
far.

\begin{figure}
    \includegraphics[width=\linewidth]{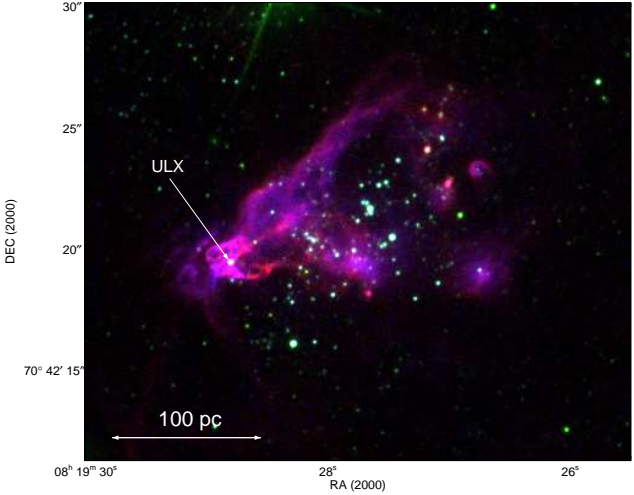}
\caption{False--colour image of the `foot nebula' HSK~70
constructed using the \HST ACS/WFC archive data (published in \citealt{hong13}). Red channel corresponds to the image obtained with F658N, green -- with F814W, blue -- with F502N filters. The ULX HoII X-1 location is shown.}\label{fig:ulx_hst}

\end{figure}

Only small part of $\sim 150$ known ULXs reveal surrounding
optical nebulae \citep*{pakull02}.
Among such objects, HoII X-1 in the Holmberg~II
galaxy -- one of the most luminous ULXs known  ($L_{bol} \sim
1.34\times10^{40} \ergs$ according to \citealt{berghea10a}). It is located in the eastern part
of the  \HII region HSK~70 (according to the catalogue by
\citealt*{hodge94}), also known as `foot nebula' owing to its
geometry (see Fig.~\ref{fig:ulx_hst}). The shape of the `foot nebula'
is most probably caused by the influence of the young star cluster
(its age is $\sim3.5 - 4.5$ Myr, according to \citealt{stewart00})
located to the west from the ULX.

A nebula with a size of about $20\times50$~pc bright in the \HeII
4686 \AA\, emission line surrounds the ULX HoII X-1
in the small eastern part (`heel') of the `foot nebula'
\citep*{pakull02, kaaret04a, lehmann05}.  \citet{moon11} revealed
that the \HeII nebula is more extended and have a size of $\sim
122$~pc. Our long-slit observations presented in \citet*{Egorov13} also revealed the \HeII emission at such distance from the ULX. These studies support the photoionization mechanism of the \HeII line excitation by reprocessing of  HoII X-1 X-ray
emission.

\citet{lehmann05} performed the integral-field unit (IFU) spectral
observations of the nebula and investigated
ionization conditions and gas kinematics in the `heel' of the
`foot nebula'. They found that the velocity dispersion in the
H$\beta$ emission line is two times higher at the position of ULX
as compared to the HSK~70 nebula as well as significant radial
velocity variations inside the \HeII region. The authors explained
complex gas kinematics by the influence of the ULX which
dynamically perturbs the surrounding ISM via jets or the accretion
disc wind.

The extended synchrotron radio source surrounds the ULX
\citep*{miller05} and coincides with the \HeII emission line
nebula in the `heel' \citep{cseh12}.  \citet{cseh14,
cseh15} discovered  a collimated jet structure
responsible for the inflation of the surrounding radio bubble.
Their results suggest that Ho~II X-1 is powered by a black hole of
a mass in the range of $25-100 M\odot$ accreting at a high
Eddington rate with intermittent radio activity.

\citet{berghea10a, berghea10b} have detected the [O~\textsc{iv}] 25.89 $\mathrm{\mu m}$ emission in the `heel' and 
concluded that the photoionization by the soft X-ray and far ultraviolet (FUV) radiation from the ULX is responsible for the [O~\textsc{iv}]  emission, while the shocks likely contribute very little.
\citet{heida14} discovered the near-infrared (NIR) counterpart;
its NIR luminosity far exceeds the expected jet
contribution. The authors related the NIR excess to a red
supergiant companion or to the nebula surrounding the ULX.

The data mentioned above suggest the influence of both the ULX and nearby young star
cluster on the surrounding ISM. The accurate
separation of the ULX contribution from the totality of processes regulating
the morphology and dynamics of the `foot nebula' is necessary.


In our recent paper \citep[][hereafter \citetalias{Egorov16}]{Egorov16} we
report the results of ionized gas structure and kinematics
investigations of the star formation complexes in Holmberg~II
galaxy. In this paper, we present the detailed study of the ULX vicinity using the 
data described  in \citetalias{Egorov16} and some  additional data sets. 

\section{Observations and data reduction}\label{sec:obs}

\begin{table}
    \caption{Log of FPI observational data}
    \label{tab:obs_data}
    \begin{tabular}{llllllll}
        \hline
        Data set  & Date of obs & $\mathrm{T_{exp}}$, s  & $\theta$, $''$ &  $\Delta\lambda$,~\AA  & $\delta\lambda$,~\AA \\
        \hline
        \Ha \#1 & 26/27 Apr 11 & $40\times240$  &    1.4  & 8.7 & {0.48  } \\
        \Ha \#2 & 16/17 Dec 14 & $40\times160$   &   1.4  & 8.7 & {0.48  } \\
        \SII   & 30/31 Oct 11 & $40\times360$  &    1.1 & 9.2  & {0.48 } \\
        \OIII   & 27/28 Feb 09 & $27\times200$  &   1.5 & 7.6 & {0.81 } \\
        \hline
    \end{tabular}

\end{table}

Observations were made at the prime focus of the 6-m telescope of
SAO RAS using a scanning Fabry--Perot interferometer (FPI) IFP751
mounted  inside the SCORPIO-2 multi-mode focal reducer
\citep{scorpio2}. A detailed description of the
observations performed and data reduction are presented in our \citetalias{Egorov16}. All
the data cubes have a field of view of $6.1\times6.1$~arcmin,
where each pixel of 0.71 arcsec in size contains a 36- or 40-channel
spectrum of a small region $\Delta\lambda$ around the desired
emission line. Table~\ref{tab:obs_data} summarizes the
properties of the observed data: name of data set, date of
observations, exposure time, final spatial resolution ($\theta$),
spectral range ($\Delta\lambda$), and spectral resolution
($\delta\lambda$ -- FWHM of the instrumental profile).

In addition to the mosaic of two observed data cubes in the \Ha
line used in \citetalias{Egorov16}, we also analyse the  data cubes
in the \SII$\lambda$6717~\AA\, and \OIII$\lambda$5007~\AA\, emission lines reduced
in the same way as the \Ha data cube. The  \SII data cube was
obtained with the same \mbox{SCORPIO-2} device, while  \OIII observations
taken from the SAO RAS archive were performed for a program by P.
Abolmasov with the focal reducer SCORPIO \citep{scorpio} and FPI having twice lower spectral
resolution (see Table~\ref{tab:obs_data}). The original \OIII data
set was completed only by 75 per cent on 36 scanning channels
because of bad weather conditions. Fortunately, our new wavelength
calibration algorithms \citep{Moiseev2015} allows us to restore
emission-line spectra in the region around the HSK 70 nebulae.

Before analysing, we have subtracted a circular rotation model for Holmberg~II from the data cubes (the procedure is described in \citetalias{Egorov16}). 
It allowed us to analyse the local ionized gas motions, not related to the regular rotation of the galaxy disc. Due to that, the velocity scale in all Figures is shown to be relative to the mean velocity.

Because of the lower signal-to-noise ratio for the \SII and \OIII
data cubes in comparison with the \Ha one, we used them only for
additional control of the results obtained
from the \Ha line. We fitted the line profiles
in each pixel of the reduced data cubes with one- or two-component
Voigt function  (a convolution of the Lorentzian and
Gaussian functions corresponding to the FPI instrumental profile
and to the broadening of the observed emission lines, respectively) to get 
information about ionized gas kinematics in the regions (the
procedure is described in \citealt{MoiseevEgorov2008}).

\section{Results and Discussion}\label{sec:dis}

\begin{figure}
    \includegraphics[width=0.95\linewidth]{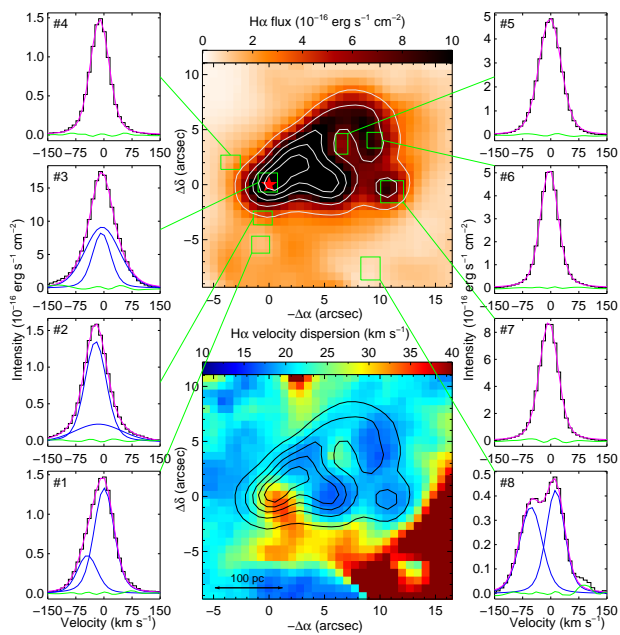}

\caption{`Foot nebula' mapped with the scanning  FPI  in the \Ha line.
The line flux (top image) and the line-of-sight
velocity dispersion (bottom image) are shown. Isophotes of the \Ha
brightness are overlaid. A red star shows the position of the ULX
in (0,0) point. Green rectangles denote the
area of the emission line profiles integrations. Black lines
denote the observed line profiles, blue lines show individual
Voigt components, magenta lines -- model spectrum, green lines -- residuals of its subtraction.
}  \label{fig:profiles}

\end{figure}

\subsection{Ionized gas kinematics of the `foot nebula'}

As we noted above, \citet{lehmann05} performed the IFU
observations of the nebula around HoII X-1 and revealed a
complexity of the `foot nebula'  kinematics.
While their observations provide much larger
spectral range, our data has 3--15 times better spectral
resolution and slightly better spatial resolution and sampling. It
allows us to decompose a line profile in the regions,
where it shows asymmetry or significant broadening.

Figure~\ref{fig:profiles} shows several examples of the \Ha line
profile which correspond to different parts of the `foot nebula'
and the results of their decomposition. The top image in this
Figure corresponds to the \Ha intensity distribution in the nebula according to our FPI data, the bottom image shows the line-of-sight velocity dispersion ($\sigma_\mathrm{H\alpha}$) map, free from instrumental broadening.

The  \Ha line profile could be fitted by a single
 narrow component ($\sigma_\mathrm{H\alpha} = 13-18
\kms$) in the most part of  HSK~70 nebula. Our data show two
regions with peaks of velocity dispersion consistent with
 \citet{lehmann05}. One of them corresponds to the ULX vicinity, while another one is located in
the north-western part of the nebula.

The line-of-sight velocity dispersion is two times larger near  the ULX, that agrees with the estimates  by \citeauthor{lehmann05} Broadening of the line profiles probably caused by the contribution of the broad component from the ULX to
the observed line-of-sight spectrum. Indeed, the \Ha line profiles
there could be fitted by narrow and broad
($\sigma_\mathrm{H\alpha} = 42-60 \kms$) components (profiles  \#2, 3 in Fig.~\ref{fig:profiles}).

The region of significant contribution of the broad component to
the emission line profile surrounds the ULX. It coincides with the
northern part of the area of elevated velocity dispersion around
the ULX (in the bright part of the HSK~70 nebula), while two resolved
narrow components are presented  in the southern part of this area
(see profiles \#1, 8). The \Ha line profile broadening
there, as well as its clear separation in two components, is
caused by the interaction of HSK~70 with the ionized supershell
south-westward of it, which origin is not related to
the `foot nebula' (see \citetalias{Egorov16}).

The \Ha line profiles of the second region of the increased velocity
dispersion inside the HSK~70 nebula  (profile \#5 in
Fig.~\ref{fig:profiles}) might be represented by a single component
with $\sigma_\mathrm{H\alpha}$ up to 24~$\kms$. The \Ha image
 reveals a shell-like structure in this area.
Its comparison with the  stars location  in Fig.~\ref{fig:ulx_hst}
allows us to conclude that there we observe an expanding shell
around a young star cluster, but our resolution is insufficient
to clearly separate the individual components corresponding to its
approaching and receding sides.

Therefore, the `foot nebula' is located at the north-eastern edge of the kinematicaly active star formation complex and is regulated by the influence of the young star cluster (which blew up the ionized superbubble around it) and of the ULX. The presence of the broad component in emission line profiles  points to the significant influence of the ULX to the surrounding ISM.


\subsection{Bow-shock around the ULX?}

\begin{figure*}
    \includegraphics[width=0.99\linewidth]{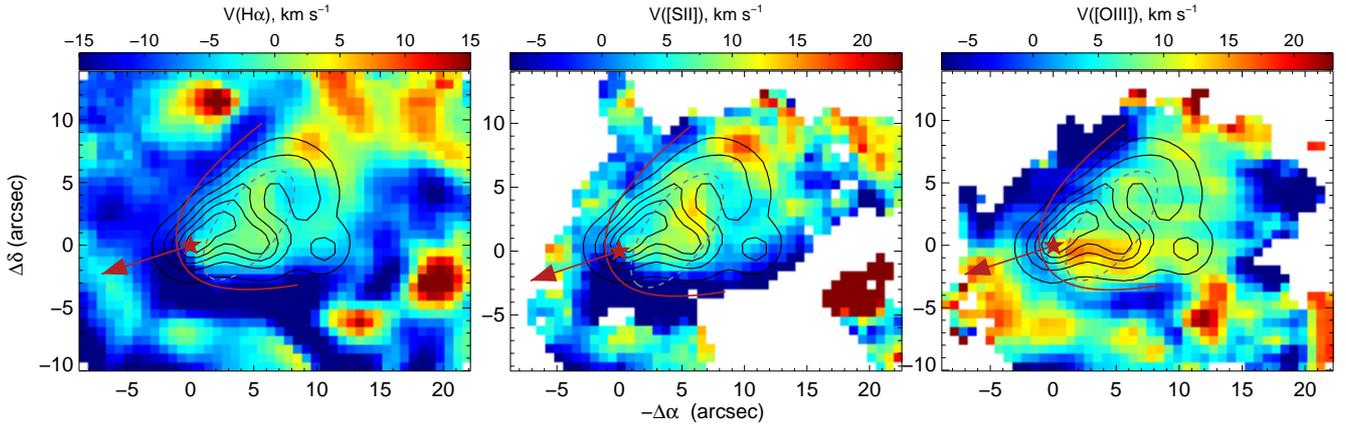}
\caption{`Foot nebula' HSK~70 and its surroundings: line-of-sight
velocity fields in the \Ha (left), \SII (middle), and \OIII
(right) emission lines. Pixels with signal-to-noise ratio less
than 5 were masked. A red star denotes the HoII X-1 position,
black contours are \Ha isophotes from the previous figure, a grey
dashed ellipse shows the borders of a young cluster. A red curve shows the modelled shape of a bow shock that
might be induced by the ULX motion in the direction shown with an arrow (see text).}\label{fig:ulx_velmaps}
\end{figure*}

We found an intriguing structure in the ionized gas line-of-sight
velocity distribution constructed by the pixel-by-pixel fitting of the
data cubes with single-component Voigt function.
Figure~\ref{fig:ulx_velmaps} shows velocity fields for the HSK~70 nebula
and its surrounding in the H$\alpha$, [S~\textsc{ii}], and \OIII emission lines.
These maps reveal blue-shifted `arc' surrounding the `foot nebula'
from the east of the ULX. This `arc' is better seen in the \Ha but is also detected in both \SII and [O~\textsc{iii}] lines. Its
line-of-sight velocities are shifted by -10 -- -15 $\kms$ from
the mean velocity in the nebula. Here we are trying to explain this observed feature.

The `arc' is located at the edge of the bright HSK~70 nebula and corresponds to faint emission of the ionized gas. Its mean brightness in \Ha line is about $4.6\times10^{-16}~\mathrm{erg~s^{-1}cm^{-2}arcsec^{-2}}$ ($2.7\times10^{-16}$ and $9.5\times10^{-16}$ in the faint north-east and bright east parts respectively). For comparison, the mean brightness of the HSK~70 nebula is $1.9\times10^{-15}~\mathrm{erg~s^{-1}cm^{-2}arcsec^{-2}}$. Due to this fact, one may suggest that we observe  some sort of the edge effects at the border of the bright \HII region and  background or foreground diffuse emission. However, 
in this case it is straightforward to expect such features at the borders of other bright nebulae in the galaxy, but only HSK~70 reveals it. Moreover,  the reasonable question arises -- why is this feature not homogeneous or randomly distributed but has a shape that clearly matches the shape of the nebula? Even if the `arc' is indeed a diffuse emission, it most probably should be regulated by something tied with the `foot nebula', and the most obvious source is the ULX.

Note that the `blue filament' at $(\Delta\alpha, \Delta\delta) = (12,-8)$ connected to the `arc' (see H$\alpha$ velocity field in Fig.~\ref{fig:ulx_velmaps})  has a different origin. Contrary to what observed in the `arc', the \Ha line profile there is clearly decomposed by  two components (see profile \#8 in Fig.~\ref{fig:profiles} as an example), that is the reason why this structure is blue-shifted in the velocity fields constructed by a fitting with a single Voigt component. The complex kinematics of the ionized gas in this region located to the south of the ULX is discussed in our \citetalias{Egorov16}. 

Both the `arc' shape, as well as
the `heel' shape of the `foot nebula' look like a bow shock.
We propose that it may be indeed a bow shock caused by the ULX
motion. Let us check now, whether the ULX escaping from the nearby young star
cluster might create such a large bow shock having the shape of
the observed `arc'.

According to the analytical solution for the shape
of a stellar wind bow shock in the thin-shell limit as derived in
\citet{wilkin96}:
\begin{equation}
\label{eq:bow_shock_shape}
R=R_0\csc\theta\sqrt{3(1-\theta\cot\theta)},
\end{equation}
where $R$ is the distance from the source at which a bow shock front
is observed in direction of an azimuthal angle $\theta$. The
stand-off radius $R_0$ can be expressed as \citep{baranov71}:
\begin{equation}
\label{eq:bow_shock_rad}
R_0=\sqrt{\frac{\dot{M_w}\upsilon_w}{4\pi\rho_\mathrm{AMB}\upsilon_\mathrm{ULX}^2}},
\end{equation}
 where $\dot{M_w}$ is the wind mass-loss rate; $\upsilon_w$ is the
terminal wind velocity; $\upsilon_\mathrm{ULX}$ is the velocity of
the ULX; and $\rho_\mathrm{AMB}$ is the ambient gas density. According to \citetalias{Egorov16},  the  ambient volume density in the area is
$n_\mathrm{H} \simeq 0.7\ \mathrm{cm^{-3}}$; that corresponds to
$\rho_\mathrm{AMB} \simeq 1.6\times10^{-24}\ \mathrm{g\ cm^{-3}}$,
if   helium  contribution in total gas mass is taken into account.

We estimate the velocity of HoII X-1 from the following
considerations. The borders of the young cluster associated with the
HSK~70 nebula are  shown in Fig.~\ref{fig:ulx_velmaps}. Most
probably the ULX was ejected from that cluster. \citet{stewart00}
have found the age of this cluster is $3.5-4.5$ Myr. For the
$\upsilon_\mathrm{ULX}$ estimation, we assume that the ULX started to move from
the cluster's centre in direction of its current position $\sim
4$ Myr ago. In that case, the length of HoII X-1 path in galactic
plane is $290$~pc \citep[for accepted disc inclination
$i=49^\circ$ according to][]{Oh11} that gives
$\upsilon_\mathrm{ULX} \simeq 70 \kms$. This value agrees with the
typical velocities of  ULXs in the Antennae galaxy
($\upsilon_\mathrm{ULX} = 80 - 90 \kms$) according to
\citet{poutanen13}.

Unfortunately, there are no HoII X-1 wind parameters known,
but we still may estimate $R_0$ using the following suggestions.
\citet{fabrika15} showed that the nature of the well known
Galactic X-ray source SS~433 is most probably the same as of the
extragalactic ULXs. Hence, in our  analysis we may use
the values of $\dot{M_w} = 5\times10^{-4} M_\odot\ yr^{-1}$
\citep*{begelman06} and  $\upsilon_w \sim 1500 \kms$
\citep{fabrika04} adopted for SS~433. On the other hand, \cite{pinto16} discovered the very high outflow velocities $\upsilon_w \sim 0.2c$ from two ULXs. Due to the fact that the supersonic wind power of ULXs is often comparable to their X-ray luminosity, $\dot{M_w} \sim 1\times10^{-5} M_\odot\ yr^{-1}$ for HoII X-1. Both estimates give almost the same value of $\dot{M_w}\upsilon_w$.

Using parameters estimated above, we found $R_0 \simeq 27$~pc from
Equation~(\ref{eq:bow_shock_rad}) and the bow shock shape in
galactic plane computed from Equation~(\ref{eq:bow_shock_shape}).
Red curves in Fig.~\ref{fig:ulx_velmaps} correspond to the shape
of a bow shock after projection to the sky plane using the
parameters of Holmberg~II galactic disc orientation from \citet{Oh11}. These curves
well coincide with the `arcs' of  negative line-of-sight
velocities  in all three emission lines. Note that variations of
galaxy inclination and ULX velocity would change these curves
insignificantly. Therefore, we suppose that the observed features in
the velocity field indeed could be treated as a sign of a bow
shock from the moving ULX HoII X-1.

The presented  results suppose that HoII X-1 was ejected
from the cluster in the galactic plane, or at least with the small
angle to that. We performed several calculations proposing the
larger angle and found that our adopted parameters give the best
agreement with the observations.

The shape of the observed `arc' is consistent with the calculated model of bow shock from the moving ULX. But it is still desirable to find some another pointing on the bow shock origin of the `arc'. Unfortunately, relatively low spatial resolution and strong contamination by the bright \HII region makes the comparison with the existent models \citep[e.g.,][]{meyer16} of the emission from bow shocks unreliable. Optical spectroscopy might provide an additional clue, 	but as it follows from Fig.~\ref{fig:ulx_velmaps}, we deal with the slow shocks, whose contribution to the emission is negligible -- according to the \cite{allen08}, the significant influence of the shocks to the spectrum starts from the velocities of about $100 \kms$. But we still may observe the local high velocity filaments in the bow shock having the elevated ratio of the sensitive to the collisional excitation lines. We indeed see the increase of the \SIIHa\ ratio at the positions of the `arc' in the spectra published in \cite{Egorov13} (and much better in the archival spectra of HoII X-1 obtained by \citealt{fabrika15} at the SUBARU telescope), yet its values are still too low to be interpreted as sign of a shock excitation. It may be explained by 
	such high velocity filaments,
	 or by the lower ionization state at the border of \HII region. In the first case it might be revealed, for example, by the high-resolution narrow band imaging with \textit{HST}. Unfortunately, Holmberg~II has never been observed by \HST in the \SII emission line.

Summing up, the `arc' indeed could be a bow shock caused by the ULX moving from the star cluster in HSK~70, yet the additional confirmation is desirable. In that case it is the first kinematic evidence of ULX's escape from its parental star clusters. Our conclusion agrees with the results obtained by
\citet{kaaret04b, poutanen13}, who showed that the majority of
ULXs are associated with young ($t < 5$ Myr) star clusters, yet
reside outside them. 

\section{Summary}\label{sec:sum}

The HSK~70 nebula (`foot nebula') related with the ULX HoII X-1 was observed in
the H$\alpha$, [S~\textsc{ii}], and \OIII emission lines with a scanning Fabry--Perot interferometer. Using these data 
we obtained the following results:

\begin{itemize}
\item Two regions of the increased line-of-sight velocity
dispersion are observed inside the nebula. Relatively high
velocity dispersion in one of them most probably corresponds to
the expanding ionized superbubble around the young star cluster,
while the second one clearly reveals the presence of a broad
component caused by the ULX influence on the ISM.


\item The velocity fields  reveal  a blue-shifted `arc', which shape  looks like a
bow shock. The  comparison with the analytical solution
by \citet{wilkin96} for the shape of a stellar wind bow shock in
the thin-shell limit  shows that this structure indeed could be a
bow shock created by the ULX moving from the central part of the
star cluster in HSK~70.

\end{itemize}

The most important result is that we found kinematic evidence of a bow shock around the ULX which in our view strongly suggests its escape from the parental star cluster.
Up to now all the conclusions about ULXs' escape were based only
on their location outside the cluster.

\section{Acknowledgements}

Authors are grateful to the anonymous referee for constructive comments and suggestions and also to S.~I.~Blinnikov and S.~N.~Fabrika for useful discussions. The work is based on
observations obtained with the SAO RAS 6-m telescope carried out with the financial support of the Ministry of
Education and Science of the Russian Federation (agreement No.
14.619.21.0004, project ID RFMEFI61914X0004). Figure~1  
in this article is based on the observations made with the
NASA/ESA Hubble Space Telescope and obtained from the Hubble
Legacy Archive which is a collaboration between the Space
Telescope Science Institute (STScI/NASA), the European Space
Agency (ST-ECF/ESAC/ESA), and the Canadian Astronomy Data Centre
(CADC/NRC/CSA). The study  was supported by the RFBR, project 14-02-00027. AVM and OVE are also
grateful for the financial support (grant MD3623.2015.2)
from the President of the Russian Federation.

\label{lastpage}

\end{document}